# Supply Chain Optimization Strategies: An Empirical Study on Fresh Product Delivery Routes

Yuan yuan[1]   Xiaoke Xie[2]   Yulei Xie[3]

1. School of Business, Shanghai Normal University Tianhua College
2. School of Management, Victoria University of Wellington
3. research institution, Huaxin Securities

**Abstract:** In recent years, with rising consumer demand, fresh products have gained increasing attention, leading to rapid growth in the fresh food market. However, due to their perishable nature and sensitivity to storage conditions, fresh products are vulnerable to damage during transportation. Improper handling, excessive transit times, and physical impacts can result in significant losses. As a result, enhancing the efficiency of fresh product distribution while maintaining quality has become critical to the further development of the fresh food industry. Using Y chain supermarket as a case study, this paper investigates the logistics of fresh product distribution, identifying current challenges and inefficiencies. Through literature review, expert interviews, and comparative analysis, the study offers strategic recommendations for optimizing fresh product delivery routes to improve distribution efficiency and product quality.

# 1. Introduction

In recent years, as consumption levels continue to rise, the demand for fresh products has grown rapidly, with consumers placing greater emphasis on quality. However, fresh products are inherently perishable, with high demand and diverse varieties, making their distribution particularly challenging. The delivery of fresh goods requires stringent controls over time, packaging, and temperature. These unique characteristics have placed fresh product distribution at the forefront of attention in the logistics industry.

Y Chain Supermarket opened its first "farm-to-supermarket" store in Fujian in 2000 and has since expanded nationwide, covering 29 provinces with over 1,000 stores. As one of the pioneers in introducing fresh products into modern retail supermarkets, Y Chain Supermarket has positioned fresh goods as a core focus of its sales strategy. Currently, its range of fresh products exceeds 3,000 items, with over 40% of each store's operational area dedicated to fresh agricultural products. The gross profit margin on these products consistently stays between 12% and 13%, outperforming the industry average of 7% to 8%. According to the 2023 annual report, the supermarket's fresh product gross profit margin reached 12.95%, contributing to 36.2% of the company's total gross profit. This is significant considering the industry benchmark for fresh product gross margins typically hovers around 6%, and in some cases, is even zero.

Y Chain Supermarket's logistics network is closely aligned with its store expansion, simplifying the complex relationships between stores and suppliers by connecting them directly to regional distribution centers. This has streamlined the supply chain, turning the coordination between stores into an internal process. The fresh product logistics process includes stock preparation, sorting, and delivery. To ensure the freshness of products, cold chain logistics play a crucial role in Y Chain Supermarket's operations.

However, in China, cold chain transportation rates for primary agricultural products have long lagged behind those of developed countries. For example, in developed markets, 80% to 90% of fruits, vegetables, meat, and seafood are transported via cold chains, while in China, the rates are only 15%, 30%, and 40%, respectively. Additionally, the rate of cold chain breaks in China is significantly higher, at 67%, 50%, and 42%, compared to developed nations.

Given these challenges, this paper focuses on optimizing the fresh product delivery routes of Y Chain Supermarket from a supply chain perspective. By improving the logistics and distribution system, the goal is to reduce product losses significantly and optimize delivery times as much as possible.

## 2. Literature Review

In recent years, with the widespread application of artificial intelligence and information technology in the logistics field, scholars have begun to focus on how to enhance logistics efficiency through technological means. Li Fan et al. (2022) pointed out that the existing employment models for fresh agricultural product delivery have not maximized their effectiveness, making it urgent to optimize human resource management. Meanwhile, Xinke Du (2020) explored the application of artificial intelligence in logistics warehousing and found that AI technology can significantly improve logistics delivery efficiency, providing new ideas for addressing the timeliness and freshness issues in fresh product delivery.

Additionally, Miao Xiaohong et al. (2020) noted that inventory management and information alignment from the perspective of the supply chain are important trends in fresh logistics management. Further research indicates that supply chain optimization based on artificial intelligence technology has become an effective way to enhance logistics efficiency. For instance, Xinke Du, Guanqing Shi, and Yilin Zhao (2023) studied the application path of artificial intelligence technology in e-commerce supply chains and pointed out that AI can help companies better cope with challenges in fresh logistics by reducing logistics costs and improving delivery efficiency.

In the field of cold chain logistics, Xie Dong (2021) proposed that China's cold chain technology development is lagging and costs are high, leading to a significant increase in logistics costs for fresh retail enterprises. To address this issue, Du X et al. (2022) suggested strategies to enhance sustainable energy production through oil price volatility and energy efficiency management, indirectly promoting a reduction in cold chain logistics costs.

Foreign research has certain advantages in the optimization of fresh logistics routes and the application of intelligent technologies. For example, Jiawei Liao (2022) used a fuzzy comprehensive evaluation method to construct a mixed-integer programming model for logistics node site selection, optimizing the freshness and loss costs of goods in the overall delivery process. Similarly, Qun Feng (2023) constructed a distribution path optimization model for fresh cold chain products considering green costs, aiming to achieve a win-win situation for environmental protection and corporate benefits through energy conservation and emission reduction.

Moreover, with technological advancements, artificial intelligence-based path optimization algorithms are increasingly being applied to practical logistics scenarios. Xinke Du (2021) applied the Floyd algorithm for route optimization in the transportation of hazardous materials, demonstrating the great potential of AI in optimizing logistics delivery routes. Likewise, Sun Jun et al. (2022) combined genetic algorithms with real cases to optimize enterprise delivery routes, further enhancing delivery efficiency.

## 3. Analysis of the Current Situation and Issues

### 3.1 Introduction to Y Chain Supermarket

Y Chain Supermarket was established in 2001, with its headquarters located in Fuzhou, Fujian Province. It is a well-known convenience supermarket among the people of Fujian. After more than a decade of development, Y Chain Supermarket has rapidly risen to prominence, successfully entering the ranks of China's top 100 enterprises. It is a leading company in both the "circulation and agricultural industrialization" sectors in Fujian Province, and as a "China Famous Trademark," it plays a significant role in people's lives.

By the end of 2010, Y Chain Supermarket went public. The company is supported by modern logistics and is backed by modern agriculture and the food industry. It has developed into a large group enterprise led by retail, with industrial development as its foundation. In recent years, the number of new stores opened by Y Chain Supermarket has surged. This rapid growth can be largely attributed to its unique fresh product business model and its timely embrace of the "Agriculture-Wholesale Supermarket Transformation."

Today, Y Chain Supermarket has become a large group company, with fresh product sales as its core focus and the main industry being fresh product processing. It is the first fresh product chain supermarket in China, introducing fresh produce into supermarkets. This has not only satisfied the daily needs of the public but has also promoted the development of the Chinese market economy, bringing significant economic benefits to the group. Currently, through continuous innovation, Y Chain Supermarket has become a publicly listed company that integrates retail and logistics. By selling and processing fresh products, the group consistently ranks among the top in economic indicators within its industry in China.

Approximately 93.12% of Y Chain Supermarket's revenue comes from merchandise sales, while the service sector has steadily increased over the past six years, rising from 4.62% in 2015 to 6.88% in 2020. This growth is primarily due to increased rental income and a higher level of service provided to suppliers. In terms of product classification, all categories have seen significant improvements, with food items (including clothing) increasing from 28.783 billion yuan in 2015 to 45.304 billion yuan in 2020, and fresh and processed products rising from 26.101 billion yuan in 2015 to 41.481 billion yuan in 2020. The number of Y Chain Supermarket stores has consistently grown at a high rate since its establishment, with an average annual increase of one to two hundred new stores over the past five years. The expansion of Y Chain Supermarket's stores has led to a corresponding growth in business scale, resulting in a continuous upward trend in revenue and profit levels.

In addition, Y Chain Supermarket began upgrading shopping experiences, optimizing product structures, and continuously promoting the construction of a transparent supply chain in 2023. However, it also faces challenges such as increasingly diverse and decentralized shopping channels and the "experientialism" commonly found in large companies.

### 3.2 Issues in the Distribution of Fresh Products at Y Chain Supermarket

(1) Inefficient Delivery: Delivery efficiency is low and delivery times are unstable, negatively affecting the quality of fresh products.
(2) Poor Quality of Delivery Personnel: The lack of specialization and execution capability among delivery personnel affects the transportation and storage conditions of fresh food.
(3) Lack of Advanced Logistics Technology: The absence of advanced logistics technologies, such as GPS and automated warehousing, leads to high logistics costs and significant loss of fresh products during transportation.

(4) Difficulty in Temperature Control: There is a lack of modern logistics facilities, such as cold chain delivery systems, which impacts the freshness and safety of fresh food.

(5) Inadequate Logistics Distribution Models: The logistics distribution models do not align with the characteristics of fresh food. For example, risks associated with direct procurement and contradictions inherent in self-operated logistics, as well as the lack of specialized logistics distribution centers.

(6) High Losses During Transportation: Weak cold storage and preservation technologies, along with a low proportion of refrigerated vehicles in the freight transport fleet, result in high losses of fresh products during transportation.

### 3.3 Causes of Unreasonable Fresh Product Distribution at Y Chain Supermarket

(1) Increased Pressure from Direct Procurement

Y Chain Supermarket benefits from direct and large-scale procurement, allowing it to buy cheaper fruits and vegetables. However, this creates high delivery costs and logistical pressures. Delays in transportation can lead to spoilage, which the supermarket must absorb. Additionally, purchasing at low prices often requires cash payments, straining cash flow.

(2) Inadequate Logistics Distribution Center System

Y Chain has established its own logistics distribution centers in Fujian to reduce logistics costs and address gaps in other supermarkets' logistics. These centers collect fresh produce and distribute it to supermarkets. However, the system is still underdeveloped, leading to supply issues for fresh agricultural products.

(3) Immature Distribution Technology

Products like seafood and meat need refrigerated transport. Y Chain has set up a temperature-controlled distribution system, but current technology cannot fully support their operations. Compared to developed countries, China's logistics technology is lagging, with insufficient emphasis on training logistics professionals, limiting growth opportunities for Y Chain.

(4) Incoherence in Transportation Resources and Storage Allocation

Y Chain ensures accurate information flow to reduce product costs but faces challenges in combining transportation and storage effectively. Inconsistencies with suppliers lead to instability in the logistics chain. The need for longer handling times for fresh products conflicts with a zero-inventory approach, indicating poor coordination in resource allocation.

(5) Shortage of Professional Talent

Y Chain lacks specialized logistics personnel, facing technical challenges in preserving fresh products during transport and using refrigeration technology. The shortage of trained cold chain professionals makes it difficult to address these issues. While some universities offer relevant courses, there is little communication or resource sharing among them, exacerbating the talent shortage.

## 4.4 Optimization of Fresh Product Distribution Routes at Y Chain Supermarket

### 4.1 Model Establishment

#### 4.1.1 Model Description

(1) Problem Assumptions

The optimization of fresh product distribution routes at Y Chain Supermarket is described as follows:

The known information includes the exact locations of the distribution center, weight limits of the delivery vehicles, and fuel consumption data. Additionally, the specific geographic locations of each store, daily demand for fresh produce, and desired delivery time windows are also considered.

The distribution tasks are centrally scheduled by headquarters, dispatching vehicles from the same distribution center to deliver the required fresh produce to each store. During the route planning process, various constraints must be taken into account, such as vehicle weight limits and time window requirements for the stores.

The objective of this research is to establish a mathematical model aimed at minimizing the total distribution costs. The total cost primarily consists of two components: first, the transportation costs incurred by each delivery vehicle during transit, which are related to the transportation distance and fuel consumption; second, the penalty costs incurred due to failure to deliver within the desired time windows.

To identify an optimal distribution route plan for Y Chain Supermarket, a genetic algorithm approach will be employed. This method effectively handles complex constraints and finds the lowest-cost solution among numerous possible routes.

(2) Conditional Assumptions

① Vehicles travel at a constant speed, disregarding the impact of weather conditions on delivery speed.

② Store demand is fixed.

③ Products demanded by stores along the same route can only be delivered by one vehicle.

④ The distances between stores are fixed.

⑤ Vehicle Type: All delivery vehicles are of the same model with identical maximum weight limits.

⑥ The total weight of all loaded products does not exceed the total weight limit of the delivery vehicles.

(3) Model Parameters

The model parameters are defined in Table 4-1 below.

Table 4-1 defines the model parameters

| parameter | Definition |
|---|---|
| $K$ | Utility Vehicles, K={1,2,3,...,k} |
| $N$ | Utility Vehicles ,i,j∈N={0,1,2...,k}, 0 show distribution center |
| $\mu$ | Vehicle mileage costs per unit distance |
| $Q$ | Maximum load limit for a single vehicle |
| $t_{ij}$ | Transit time from Store i to Store j |
| $L$ | The maximum distance a vehicle can travel at a time |
| $E_i$ | The earliest service time that store i can accept |
| $L_i$ | The latest service hours that store i can accept |
| $d_{ij}$ | The distance between store i and store j |
| $a_{ik}$ | The time when vehicle k arrives at store i |
| $u_{ik}$ | Unloading time for vehicle k to arrive at store i |

| $M_1$ | A penalty cost per unit of time the vehicle arrives at the store early |
|---|---|
| $M_2$ | Penalty cost per unit of time for vehicle delay in arriving at the store |

The decision variables of the model are shown in the following equations (4-1) and (4-2).

$$x_{ijk} = \begin{cases} 1, \text{indicates that the vehicles K successively serves store i and store j;} \\ 0, \text{others;} \end{cases} \quad (4\text{-}1)$$

$$y_{ik} = \begin{cases} 1, \text{Indicates that service store i is served by vehicles K;} \\ 0, \text{others;} \end{cases} \quad (4\text{-}2)$$

### 4.1.2 Objective Function

**(1) Transportation Cost**

The transportation cost refers to the fuel consumption, tolls, and other related expenses incurred by vehicles during the delivery of goods to a specific store. In this study, the transportation cost is considered to be related to the distance traveled by the vehicle, increasing with the length of the delivery route. Here, μ\muμ represents the delivery cost per unit distance for the delivery vehicle, which is set at 1.8 CNY/km. The formula is as follows (4-3):

$$C_1 = \mu \times \sum_{i=0}^{N} \sum_{j=1}^{N} \sum_{k=1}^{K} d_{ij} x_{ijk} \quad (4\text{-}3)$$

**(2) Penalty Cost**

In the optimization problem of delivery routes for Y Fresh Chain Supermarkets, the penalty cost refers to the fees incurred when delivery vehicles fail to arrive at stores within the expected time window. This time window constraint is crucial in actual operations, as it directly impacts customer satisfaction and the operating costs of the supermarket.

Time window issues are primarily divided into two types: "soft time windows" and "hard time windows." Hard time windows require vehicles to arrive at the specified time without any deviation, which is often difficult to achieve in reality due to various uncertainties in operations. Therefore, this study adopts a vehicle routing optimization problem with soft time window constraints, which is more aligned with real operational conditions.

As shown in Figure 4-1, under soft time window constraints, vehicles are allowed to arrive at each store outside the acceptable time window, but a penalty cost must be paid. Each store has a satisfactory delivery time window [Ei,Li], where Eiindicates the earliest acceptable delivery start time for the store, and Li indicates the latest acceptable delivery end time. If the vehicle arrives within this time window, no penalty cost is incurred; however, if the vehicle arrives early or late, a corresponding penalty cost M is incurred based on the extent of deviation from the time window.

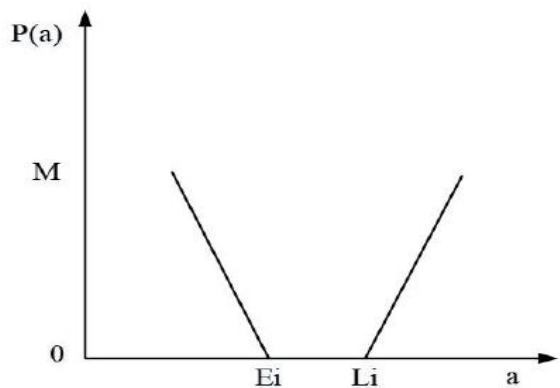

Figure 4-1: Diagram of Time Component Costs

$$C_2 = \sum_{j=1}^{N} \sum_{k=1}^{K} \{M_1 \times max[(E_i - a_{ik}), 0] + M_2 \times max[(a_{ik} - L_i), 0]\} \quad (4\text{-}4)$$

(3) Objective Function Model

The objective function model for minimizing the total cost of fresh product delivery with time window constraints is established as shown in Equation (4-5).

Equation (4-5) represents the minimum value of the sum of transportation costs and time penalty costs within this objective function model.

Constraint (4-6) indicates that each store will be visited by the vehicle only once:

$$\sum_{k=1}^{K} y_{ik} = 1 \quad i \in N \quad (4\text{-}6)$$

Constraint (4-7) indicates that after completing the delivery task for the first store, the vehicle will proceed directly to the second store for delivery.

$$\sum_{i=0}^{N} \sum_{j=1}^{N} x_{ijk} - \sum_{i=0}^{N} \sum_{j=1}^{N} x_{i(j-1)k} = 1 \quad k \in N \quad (4\text{-}7)$$

Constraint (4-8) states that each vehicle, after completing the delivery of fresh fruits and vegetables to all designated stores, must return to the distribution center.

$$\sum_{j=1}^{N} x_{ijk} = \sum_{j=1}^{N} x_{jik}, i = 0 \quad k \in N \quad (4\text{-}8)$$

Constraint (4-9) indicates that the total delivery distance for each vehicle is fixed, ensuring that the vehicle does not run out of fuel during the delivery tasks, which could disrupt the delivery process.

$$\sum_{i=0}^{N}\sum_{j=1}^{N} d_{ij}x_{ijk} \leq L \quad k \in N \quad (4\text{-}9)$$

Constraint (4-10) represents the temporal relationship between the arrival times at two consecutive stores during the vehicle's transportation process. This constraint ensures that the delivery sequence respects the time windows established for each store, preventing any overlap or violations of the scheduled delivery times.

$$a_{jk} \geq (a_{ik} + u_{ik} + t_{ij}) \times x_{ijk} \quad i,j \in N, k \in K \quad (4\text{-}10)$$

### 4.1.3 Steps of the Genetic Algorithm

(1) Encoding: Use natural number encoding to construct chromosomes that represent feasible vehicle routes.

（2）Control Parameters: Set control parameters, including crossover rate $P_C=0.7$, mutation rate P=0.1, and population size n=10.

（3）Initializtion，Initialize d=0，and randomly generate the initial population P(0)，which consists of n chromosomes，with each chromosome representing a vehicle route.

（4）Iteration: Set i=1；

（5）Route Length Calculation: Decode the i-th chromosome in the population P(d) to determine the route length.；

（6）Fitness Calculation: Calculate the fitness of each chromosome based on the total cost (transportation cost and penalty cost).

(7)Termination Check: If the algorithm's termination conditions are met, stop; otherwise, continue.

（8）Increment Index:i=i+1；

（9）Next step Check: if i<=n，returen to step 5，otherwise, proceed to Step 10；

（10）Genetic Operations: Perform maximum retention crossover, bit-based mutation, and inversion operations on the population.

（11）Generation Increment:d=d+1；

（12）Termination Check: If the termination conditions are met, stop; otherwise, return to Step 4.

## 4.2 Vehicle Route Optimization Solution

This study focuses on the distribution of Y Supermarket Co., Ltd. in Shenyang, Liaoning Province, specifically addressing the delivery to eight chain supermarket stores. A delivery route optimization model has been established for this purpose. Currently, the distribution center operates fixed delivery routes, with vehicles delivering goods along predetermined paths each day. At present, there are two fixed delivery routes in operation, as illustrated in Figure 4-4. The spatial relationship between the distribution center and the various stores is depicted in Figure 4-2.

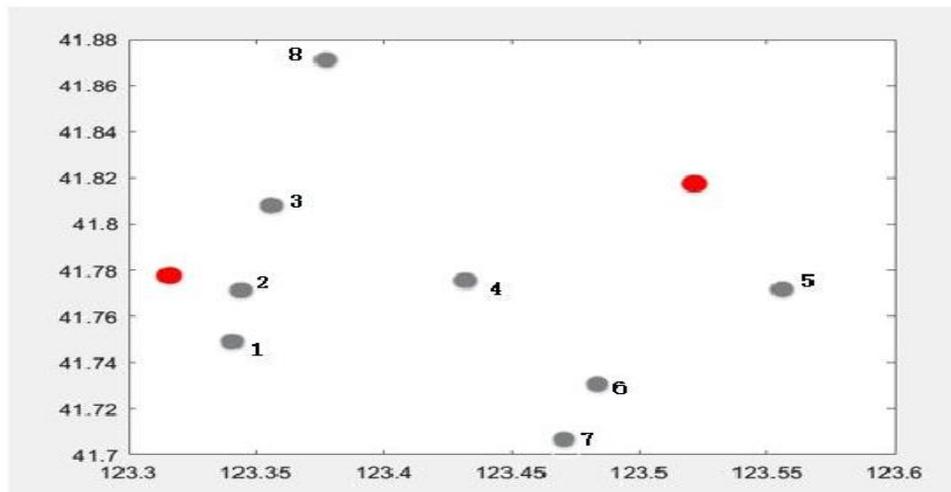

Figure 4-2 Distribution of each store (Data source :Y chain supermarket internal publication)

The fresh demand, expected time window, and loading and unloading time of each store are shown in Table 4-2.

Table 4-2 Fresh demand information Table

| Store No. | Fresh demand (tons) | Handling time (min) | Acceptable time window | Expected time window |
|---|---|---|---|---|
| 1 | 0.3 | 18 | 6:00-10:30 | 7:30-9:00 |
| 2 | 0.5 | 30 | 7:00-13:30 | 8:00-11:30 |
| 3 | 0.2 | 12 | 6:30-12:30 | 8:00-10:30 |
| 4 | 0.6 | 36 | 6:30-13:30 | 7:30-11:30 |
| 5 | 0.4 | 24 | 6:00-12:30 | 8:30-10:30 |
| 6 | 0.9 | 54 | 6:00-12:30 | 6:30-10:30 |
| 7 | 0.3 | 18 | 6:00-12:30 | 7:30-11:00 |

| 8 | 0.2 | 12 | 7:00-12:00 | 8:00-11:00 |

In this paper, a real-valued encoding method is chosen to represent the optimization scheme for the delivery route. Specifically, a sequence of numbers is used to represent the delivery path, where 0 represents the distribution center, and 1 to 8 represent the eight stores. To ensure that each chromosome represents a valid delivery path, the beginning and end of the chromosome are set to 0, indicating that the vehicle departs from the distribution center and returns after completing all delivery tasks.

Table 4-3 lists of the distance between stores（Unit: km, data source :Y chain supermarket internal publications）

|   | 1 | 2 | 3 | 4 | 5 | 6 | 7 | 8 |
|---|---|---|---|---|---|---|---|---|
| 1 | 0 |   |   |   |   |   |   |   |
| 2 | 1 | 0 |   |   |   |   |   |   |
| 3 | 4 | 2 | 0 |   |   |   |   |   |
| 4 | 10 | 10 | 9 | 0 |   |   |   |   |
| 5 | 24 | 23 | 22 | 14 | 0 |   |   |   |
| 6 | 16 | 16 | 15 | 6 | 8 | 0 |   |   |
| 7 | 15 | 15 | 14 | 6 | 10 | 2 | 0 |   |
| 8 | 9 | 7 | 5 | 8 | 21 | 15 | 14 | 0 |

Considering that there are eight stores to be serviced by two vehicles, a complete permutation of all stores is generated first. For example, given a chromosome encoded as 02145078630, this chromosome represents two specific delivery paths:

Path 1: 0 (Distribution Center) - 2 (Store 2) - 1 (Store 1) - 4 (Store 4) - 5 (Store 5) - 0 (Return to Distribution Center)

Path 2: 0 (Distribution Center) - 7 (Store 7) - 8 (Store 8) - 6 (Store 6) - 3 (Store 3) - 0 (Return to Distribution Center)

To evaluate different delivery path schemes, this paper uses a fitness function for assessment. The higher the fitness value, the better the delivery scheme. Since the optimization goal is to minimize delivery costs, an appropriate fitness function is selected to reflect this objective. In this case, the inverse of the objective function is chosen as the fitness function. This way, when the objective function value (i.e., delivery cost) is smaller, its inverse (i.e., fitness value) will be larger, indicating that the scheme is better.

Through this approach, this paper can use genetic algorithms to search for and optimize delivery paths, finding the scheme with the lowest cost and shortest distance.

## 4.3 The result of optimized distribution route

This paper uses MATLAB to program and solve the optimization algorithm. After 100 iterations, the optimized delivery route scheme is obtained. The fresh delivery routes of Y Fresh Chain Supermarket are now optimized, and a comparison is made between the logistics transport vehicle loading conditions before and after optimization. The analysis includes vehicle transportation costs, time penalty costs, vehicle loading conditions, and delivery routes.

As shown in Figure 4-3, after 7 iterations, the optimization of the fresh delivery route tends to stabilize, reaching the optimal solution state.

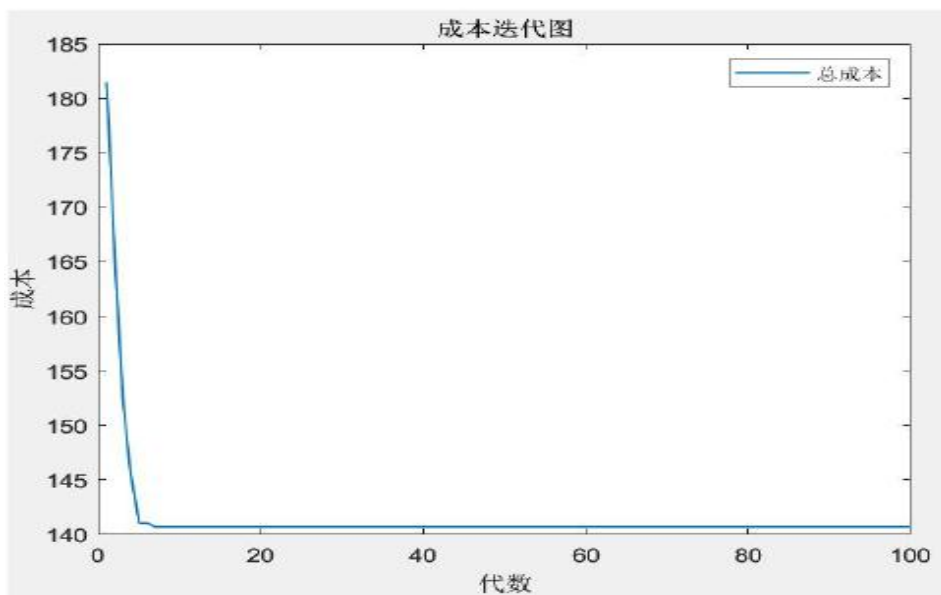

Figure 4-3 Iteration of fresh delivery path optimization

The delivery route planning for fresh fruit before optimization is shown in Figure 4-4. The vehicle delivery arrangement is as follows: Vehicle 1 is responsible for delivering to stores 8, 3, 2, and 1, while Vehicle 2 is responsible for delivering to stores 4, 6, 7, and 5. The rated load capacity of the delivery vehicles is 2 tons, and the average driving speed of the vehicles is 60 km/h.

The delivery route planning for fresh produce after optimization is shown in Figure 4-5. After optimization, the vehicle delivery arrangement is as follows: Vehicle 1 is responsible for delivering to stores 4, 1, 2, 3, and 8 in sequence, while Vehicle 2 is responsible for delivering to stores 5, 7, and 6 in sequence.

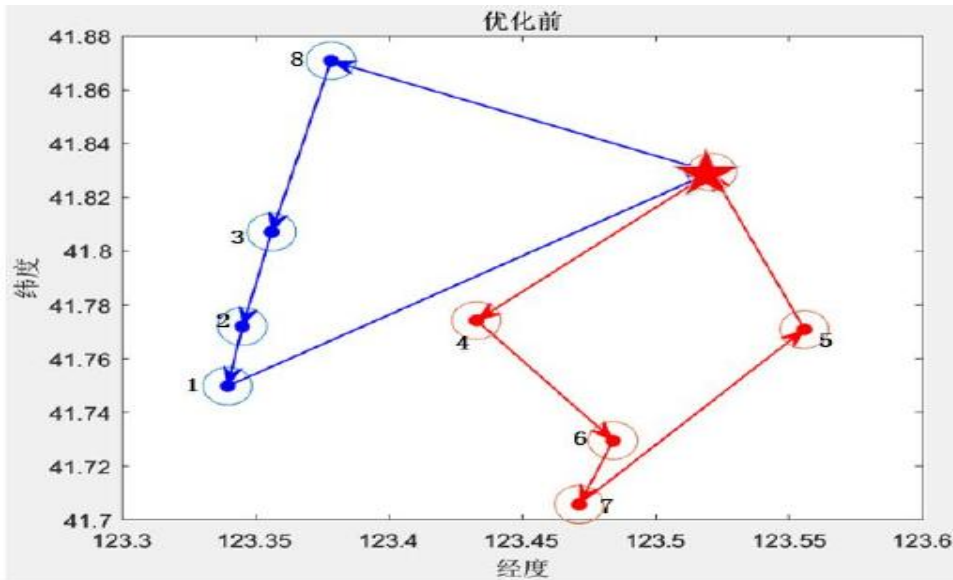

Figure 4-4 Route planning before fresh distribution route optimization

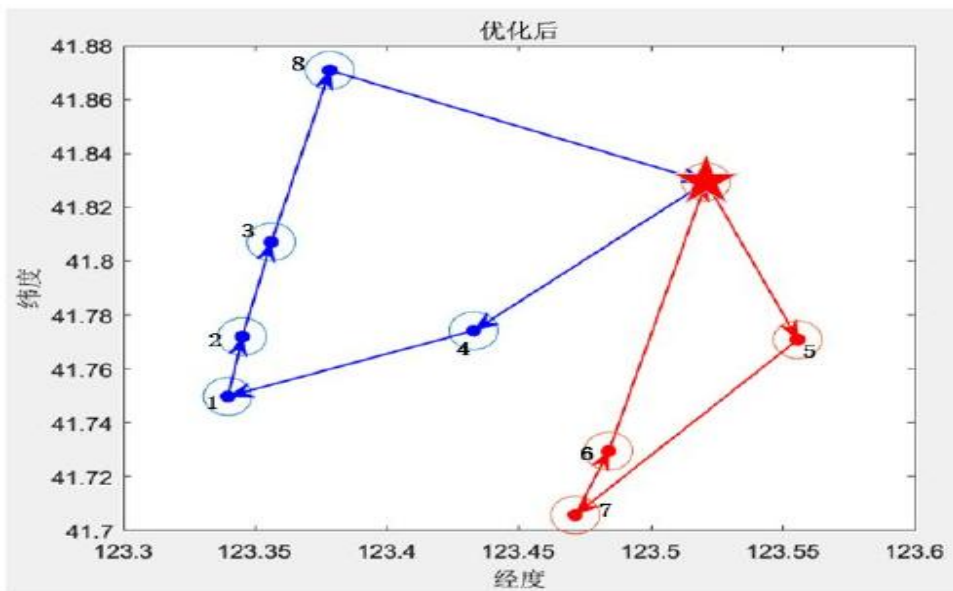

Figure 4-5 Route planning after fresh distribution route optimization

Table 4-4 Cost comparison table before and after optimization

| Index | Vehicle 1 Before | Vehicle 1 After | Vehicle 2 Before | Vehicle 2 After |
|---|---|---|---|---|
| Load Factor（%） | 60% | 90% | 110% | 80% |

| transportation cost（元） | 110 | 98 | 558 | 76 |
|---|---|---|---|---|
| penalty cost（元） | 100.5 | 0 | 0 | 0 |
| Shipping Length（min） | 112 | 78 | 104 | 121 |
| Total mileage (km) | 121 | 117 | 97 | 86 |

By comparing the cost data before and after the optimization of fresh produce delivery in Table 4-4, significant changes in delivery efficiency and cost structure can be clearly observed. Before optimization, Vehicle 1 had a loading rate of only 60%, while Vehicle 2 was overloaded with a loading rate as high as 110%. This situation required Vehicle 2 to make a second delivery, resulting in an additional transportation cost of 482 yuan. Furthermore, due to deviations from the delivery time window, a penalty cost of 100.5 yuan was incurred. The total distance traveled was 218 kilometers.

After optimization, the loading rate of Vehicle 1 increased to 90%, while the loading rate of Vehicle 2 decreased to a reasonable 80%, thus avoiding overload and the need for a second delivery. This improvement not only significantly reduced the total vehicle transportation cost from 668 yuan to 174 yuan but also eliminated the penalty cost, bringing it down to 0 yuan. Additionally, the total transportation time was shortened by 17 minutes, and the total distance traveled was reduced to 203 kilometers, saving 15 kilometers in distance and significantly enhancing delivery efficiency.

# 5. Conclusion

Y Chain Supermarket is a large retail establishment primarily focused on selling fresh agricultural products. With its unique business model that combines self-operated and direct procurement strategies, the supermarket has experienced rapid growth in its targeted market and has become well-known among consumers. This success has led to its inclusion in the "Top Ten Innovative Retail Cases in China."

As consumer demand for fresh products continues to rise, along with increasing quality expectations, this paper investigates the logistical challenges faced by Y Chain Supermarket in the distribution of fresh products. Key issues identified include low delivery efficiency, inadequate quality of delivery personnel, limited application of advanced logistics technologies, challenges in maintaining optimal temperatures for fresh foods, distribution models that do not align with the specific needs of fresh products, and significant losses during transportation.

To address these issues, this study proposes targeted optimization strategies and solutions. By employing a combination of literature analysis, qualitative and quantitative assessments, and genetic algorithms, the research successfully optimizes the delivery routes for fresh products at Y Chain Supermarket.

The optimization strategies and solutions outlined in this paper are tailored to the specific operational conditions of Y Chain Supermarket. A comprehensive investigation and analysis of its fresh delivery routes have been conducted. However, it is important to note that the models and methods described may not be directly applicable to other supermarkets of a similar nature without further analysis tailored to their unique operational contexts.